\documentclass[fleqn,usenatbib]{mnras}

\usepackage{newtxtext,newtxmath}

\usepackage[T1]{fontenc}
\usepackage{color}

\DeclareRobustCommand{\VAN}[3]{#2}
\let\VANthebibliography\thebibliography
\def\thebibliography{\DeclareRobustCommand{\VAN}[3]{##3}\VANthebibliography}

\usepackage{graphicx}	
\usepackage{amsmath}	

\title[Polarimetry of the May 2022 Lunar Eclipse]{Optical Polarimetry of the May 2022 Lunar Eclipse }

\author[I. A. Steele et al.]{
Iain A Steele,$^{1}$\thanks{E-mail: i.a.steele@ljmu.ac.uk (IAS)}
Klaas Wiersema,$^{2,3}$
Callum McCall,$^1$
Andrew Newsam,$^1$
and Manisha Shrestha$^{1,4}$
\\
$^{1}$Astrophysics Research Institute, Liverpool John Moores University, 146 Brownlow Hill, Liverpool L3 5RF, UK\\
$^{2}$Physics Department, Lancaster University, Lancaster, LA1 4YB, UK\\
$^{3}$School of Physics and Astronomy, University of Leicester, University Road, 
Leicester, LE1 7RH, UK  \\
$^{4}$Steward Observatory, University of Arizona, 933 North Cherry Avenue, Tucson, AZ 85721-0065, USA \\
}

\date{Accepted 2022 September 27. Received 2022 September 22; in original form 2022 August 16}

\pubyear{2022}

\begin{document}
\label{firstpage}
\pagerange{\pageref{firstpage}--\pageref{lastpage}}
\maketitle

\begin{abstract}
The sunlight reflected from the Moon during a total lunar eclipse has been transmitted through the Earth's atmosphere on the way to the Moon.  The combination of multiple scattering and inhomogeneous atmospheric characteristics during that transmission can potentially polarize that light.  A similar (although much smaller) effect should also be observable from the atmosphere of a transiting exoplanet.  We present the results of polarization observations during the first 15 minutes of totality of the lunar eclipse of 2022 May 16.  We find degrees of polarization of $2.1\pm0.4$\% in $B$, $1.2\pm0.3$\% in $V$, $0.5\pm0.2$\% in $R$ and $0.2\pm0.2$\% in $I$. Our polarization values lie in the middle of the range of those reported for previous eclipses, providing further evidence that the induced polarization can change from event to event.  We found no significant polarization difference ($<0.02$ percent) between a region of dark Mare and nearby bright uplands or between the lunar limb and regions closer to the disk centre due to the different angle of incidence.  This further strengthens the interpretation of the polarization's origin being due to scattering in the Earth's atmosphere rather than by the lunar regolith.
\end{abstract}

\begin{keywords}
techniques: polarimetric -- Moon -- eclipses
\end{keywords}



\section{Introduction}
\label{sec:intro}
The polarization of the reflected light from the Moon during a total lunar eclipse can act as a probe of the effect of the Earth's atmosphere on the transmission of light from the Sun.  \cite{tak2017} show that a double scattering in the Earths atmosphere combined with some form of large scale atmospheric inhomogeneity (for example due to latitudinal temperature variability) can in theory produce linear polarization\footnote{Hereafter we will use the word polarization to mean linear polarization.  Circular polarization is not considered in this paper.} fractions of a few percent.  Since the situation in a lunar eclipse is analogous to an exoplanet transit in front of its host star, we might also expect a polarization signal associated with the exoplanet atmosphere.  Such a signal would be diluted by an enormous factor \citep{tak2017} due to the non-occulted light from the primary, but may still be potentially detectable if sufficiently high precision could be obtained. As well as increasing our understanding of the polarization properties of the high levels of the Earth's atmosphere, observations of the Moon during eclipse can therefore help in the planning of future transiting exoplanet atmosphere polarization measurements.  Such measurements could potentially provide insight into the spatial distribution of large scale exoplanet atmospheric features.

Observational measurements of the polarization of the lunar disk during eclipse are rare. Integrated disk measurements using a pair of small telescopes equipped with Wollaston prisms and photo-multiplier tubes were performed by \cite{coyne}.  They found a polarization of 2.4 percent using a filter with a central wavelength at 534-nm (i.e. similar to the $V$-band) for an eclipse that occurred in April 1968.  However for observations of the the March 1960 and October 1968 eclipses they report no significant detection of polarization.

The next measurements were reported by \cite{tak2017} who used a spectro-polarimetric approach with a series of small slits sampling either the nearby sky or the eclipsed disk. They present polarization spectra and also convert their measurements to $V$ and $R$ band equivalents for the eclipse of April 2015.  They measure a degree of polarization of $2.5$ percent in $V$ and $<1$ percent in $R$.  \cite{tak2019} followed this up with an analysis of imaging polarimetric measurements of the earlier October 2014 eclipse taken by two independent telescopes.  They found the polarization was less than 1 percent in both the $V$ and $R$ bands on that occasion.  They propose the difference between this and the April 2015 measurement may be due to a different distribution of high clouds in the Earth's atmosphere between the two eclipses.

Finally \cite{strassmeier} presented the results of spectro-polarimetric observations of the January 2019 lunar eclipse in the wavelength range $742-906$-nm.  Apart from a marginal detection around the O2 band, they found no evidence of polarization at $<0.2-0.4$ percent.

From the above summary is is clear that there is still  uncertainty about the strength and variability of the polarization signal from the eclipsed Moon.  Opportunities for observing lunar eclipses from sites equipped with astronomical polarimeters are fairly rare.  Given this, we decided to make a set of multi-band optical observations of the eclipse of 2022 May 16 that was visible from the site of the Liverpool Telescope \citep{lt} at the Observatorio del Roque de Los Muchachos, La Palma.  This paper presents the results of those observations.

\section{Observations}

\begin{figure*}
	\includegraphics[width=2.0\columnwidth]{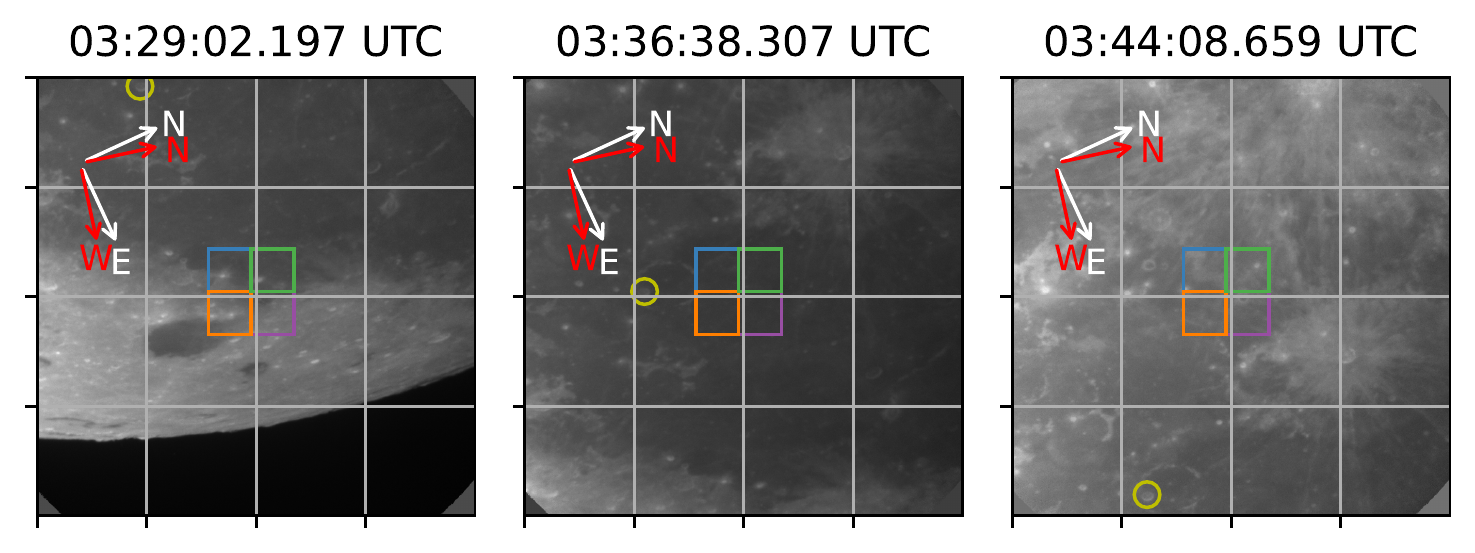}
    \caption{Full images of the eclipsed Moon from Camera 1 at the start of each $R$-band observing sequence. The four square regions outlined at centre of the images are used for polarization and photometric measurements.  The yellow circle identifies a common feature in each image.  Since the telescope was tracking at sidereal rate, the image of the moon is slowly drifting across the detector.  By the end of the observing run, the entire field of view had been traversed.  The white grid lines indicate 256 pixels (107.5 arcsec).  White arrows indicate the orientation of the standard equatorial sky coordinate system and red arrows the selenographic system which was offset by $12.5^\circ$ at the time of observation. }
    \label{fig:moon-long}
\end{figure*}

\begin{table*}
	\begin{center}
	\caption{MOPTOP filter characteristics and measured instrumental polarization ($q_0$ and $u_0$). $K$ is the offset angle between the measured EVPA and the true value and $D$ the fractional instrumental depolarization.}
	\label{tab:filters}
	\begin{tabular}{cccccc} 
		\hline
		Filter & Wavelength & $q_0$ & $u_0$ & $K$ & $D$\\
		       & (nm) & (\%) & (\%) & ($^\circ$) & \\
		\hline
		$B$ & $380-520$ &   $0.12\pm0.02$ & $-1.19\pm0.03$ & $124.7\pm0.5$ & $0.86\pm0.02$\\
		$V$ & $490-570$ &   $0.56\pm0.02$ & $-2.33\pm0.02$ & $122.8\pm0.2$ & $0.87\pm0.01$\\
		$R$ & $580-695$ &   $1.07\pm0.07$ & $-3.08\pm0.05$ & $124.1\pm0.2$ & $0.91\pm0.01$\\
		$I$ & $695-830^*$ & $1.17\pm0.02$ & $-3.38\pm0.02$ & $124.1\pm0.2$ & $0.81\pm0.01$\\
		\hline
	\end{tabular}
	\end{center}
	$^*$The $I$ band long wavelength cutoff is defined by the detector quantum efficiency and is therefore poorly defined.  
\end{table*}

Our observations used the recently commissioned MOPTOP polarimeter \citep{jermak2016, jermak2018} mounted at a folded side port on the 2.0-m Liverpool Telescope (LT).  LT is a fully robotic telescope, making it well suited for observations of time critical phenomena such as eclipses \citep{rashman} which can be executed without significant disruption to other programmes.  MOPTOP provides a field of view of $7\times7$-arcmin and uses a dual beam, dual camera design to achieve high polarimetric accuracy without overlapping images on a single camera.  The two camera fields of view overlap with a small offset of 14-arcsec due to assembly tolerances.  The instrument is equipped with a filter wheel holding $B$, $V$, $R$ and $I$ band filters (Table \ref{tab:filters}) and when operated in its default $2\times2$ binned mode has an effective plate scale of 0.42 arcsec/pixel.

The basic concept of MOPTOP is to use a continuously rotating halfwave plate and beam splitting prism to record a sequence of alternating ordinary and extraordinary images on the cameras every 22.5$^\circ$ of rotation.  Combining photometry (counts) from 4 images allows calculation of the linear Stokes parameters $q = Q/I$ and $u = U/I$ following the procedure described in \cite{shrestha}. To avoid the risk of saturation, the camera was operated in {\tt FAST} mode. 
In this mode, a full, 360 degree, rotation of the wave plate takes 8 seconds to execute, during which 16 image pairs are obtained.  This means that the frame interval is 0.5 seconds, with an exposure time per image of 0.4 seconds.  The four pairs of $q$ and $u$ values so derived are then averaged to reduce scatter introduced by slight variations in the wave plate (see \citealt{Wiersema} for an analysis of the data quality of {\tt FAST} mode MOPTOP observations). 
For these observations we chose to use ten rotations of the waveplate with a given filter before moving on to the next filter.  We shall refer to this set of ten rotations as a single "observing sequence", which takes 80 seconds to execute and generates 10 values of $q$ and $u$ for subsequent analysis.

The initial telescope pointing was calculated using the JPL Horizons ephemeris for the location and altitude of the Liverpool Telescope.  A region was chosen centred on the small crater Riccioli (selenographic coordinates 3$^\circ$S, 75$^\circ$W; \citealt{Moonbook}).  This also has a nearby region of Mare (Grimaldi). This pointing was on the opposite side of the lunar disk to the point of second contact (i.e. where the umbral eclipse began) and so in the darkest region of eclipse at the start of totality.  The region was also sufficiently close to the edge of the moon that a sky subtraction region would also be available.  Totality began on 2022 May 16 at 03:29:03 UTC and observations were successfully scheduled to begin that time (Table \ref{tab:log}) when the altitude of the Moon viewed from La Palma was $\sim 30^\circ$.  A series of observing sequences were repeated twice in the order $R-B-V-I$ and then a final $R$-band sequence was obtained.  The final observation was at 03:45:28 UTC by which time the altitude had decreased to around 28$^\circ$ and was unfortunately approaching the telescope altitude limit.  Maximum eclipse occurred later at 04:11:28 UTC and the end of totality at 04:53:55 UTC.

Although the Liverpool Telescope has the ability to track at a non-sidereal rate, it unfortunately transpired that this was not possible via the robotic control software for an object as rapidly moving as the Moon.  The result is that the image of the Moon drifts across the detector over the course of the observations (Fig. \ref{fig:moon-long}), with a total drift over 15 minutes approximately equal in size to the detector field of view.  With retrospect it is clear that we should have requested a pointing reset to the current ephemeris coordinates at the start of each observing sequence to keep the pointing at least approximately consistent.  We will discuss the implications of this image drift in Sections \ref{sec:sky} and \ref{sec:results}.

\section{Data Reduction}

\begin{figure*}
	\includegraphics[width=2.0\columnwidth]{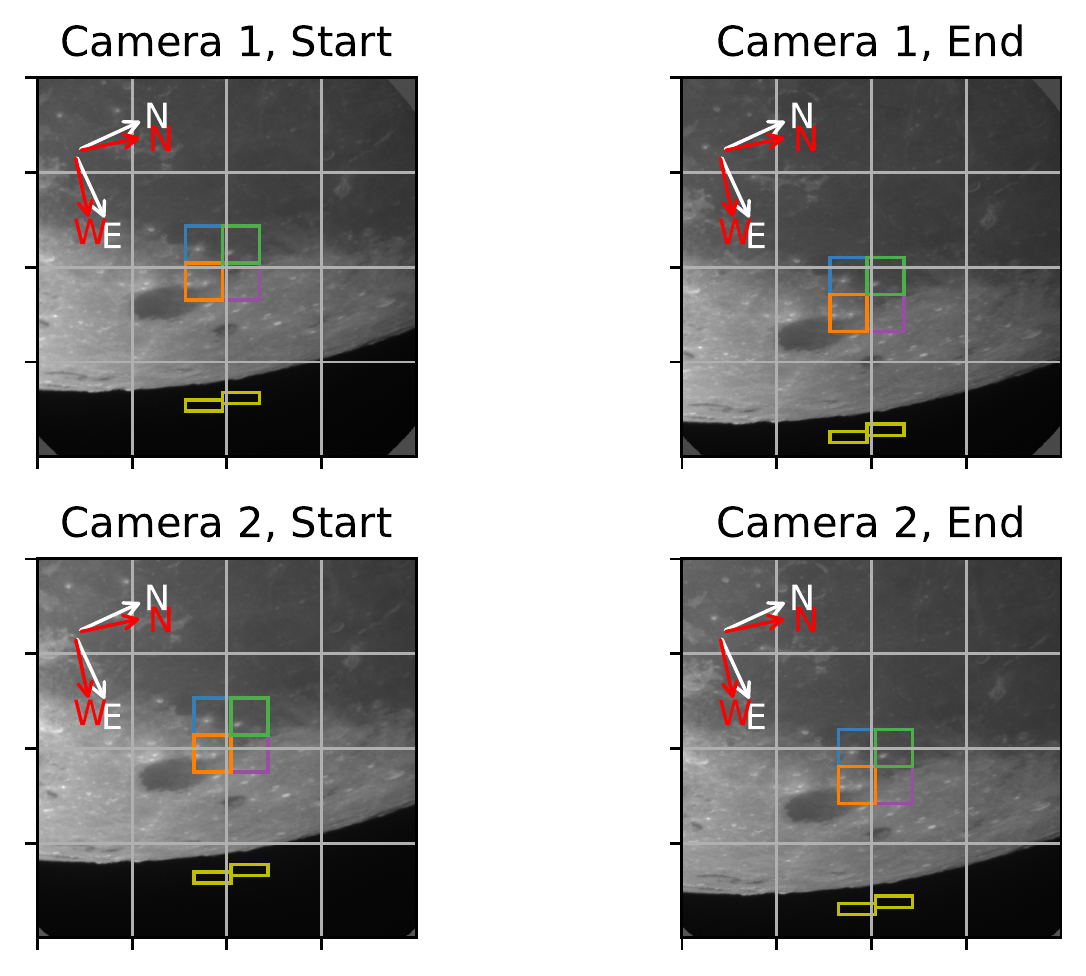}
    \caption{Images of the eclipsed Moon from the start and end of the first $R$-band observing sequence as observed with both MOPTOP cameras.  There is a small constant image offset between camera 1 and camera 2, and a drift of the images over the 80 second observing sequence due to the sidereal tracking rate.  White grid lines are drawn to make it easier to see the movement of the Moon over the detector in this interval.   The central measurement region squares are coloured to match the colour coding in Fig. \ref{fig:drift}.  Yellow rectangles the areas used for sky subtraction.  Within an observing sequence the location of these areas is adjusted to track the lunar motion. }
    \label{fig:moon-sky}
\end{figure*}

\subsection{Basic Procedure}
\label{sec:procedure}

All data from MOPTOP is dark subtracted and flat-fielded by a data reduction pipeline running at the telescope.  The dark frames used in calibration are updated daily whereas the flat fields are sufficiently stable to only require updating every few months.
We used these reduced data products in our analysis.  MOPTOP images have a slight field dependence of instrumental polarization which is more pronounced towards the edge of the images.  To avoid this, we carried out our primary analysis only considering a $200\times200$ pixel region located near the centre of the detectors.  In order to allow an empirical evaluation of errors due to field dependence (Section \ref{sec:results}), this was split into four $100\times100$ square measurement regions (Fig. \ref{fig:moon-long}), and the counts within each square averaged.  The counts from the set of 16 images per camera (a complete waveplate rotation) are then used in the equations defined by \cite{shrestha} to derive the Stokes parameters.

A nightly set of polarized and non-polarized standards \citep{schmidt,stan2,stan3} are automatically observed by LT.  The non-polarized standards are used to measure the instrumental polarization introduced by the telescope and instruments optics which is dominated by the $45^\circ$ fold mirror that feeds the instrument.  These standards are generally imaged close to the centre of the array and may therefore be used to correct the measured Stokes parameters for our lunar observations.   The instrumental polarization is characterized by the Stokes parameters $q_0$ and $u_0$ which must be subtracted from measurements of science targets to remove the effect.  The standard star observations were analysed using a MOPTOP pipeline designed for polarimetry of point sources which caries out aperture photometry on the images and applies the procedure outlined in \cite{shrestha} to calculate $q$ and $u$ values and associated errors.  In Table \ref{tab:filters} we list the results of this analysis for all standard stars observed between 2022 April 13 and 2022 July 31.  We note no systematic effects with time were apparent.\footnote{These values differ from those reported in \cite{shrestha} as new cameras were installed into the instrument in March 2022.} 

The polarized standards observed over the same time period were used to measure the affect of instrumental depolarization in MOPTOP by comparison with the catalogue values.  The resulting per-filter fraction depolarization ($D$ - Table \ref{tab:filters}) can then be be divided into the measured degree of polarization of science targets.  The polarized standards also allowed calibration of the angle ($K$) between the telescope Cassegrain rotator and the Electric Vector Polarization Angle (EVPA) following the method outlined in \cite{blazars}.

We also carried out photometry of our frames in the measuring regions (Fig \ref{fig:moon-long}).  This was calibrated by comparison with a standard star observed at the same airmass following the eclipse.  The measured mean magnitudes in each region were then combined by calculating the mean and standard deviation over the four regions and converted to magnitudes per square arcsecond.   The results of this analysis are listed in the final column of Table \ref{tab:log}.
\subsection{Sky Subtraction}
\label{sec:sky}

\cite{tak2017} discuss the difficultly of accurate sky subtraction for lunar eclipse data.  There is of course no way to measure the sky background (which might be more accurately termed foreground) directly in front of the lunar disk.  The principle source of illumination of the foreground will be scattered light from the Moon itself either in the Earth's atmosphere or the telescope/instrument optics.  \cite{tak2017} showed that the scattered light near the moon during their observations had a similar spectral energy and polarization distribution to their measurement of the lunar disk, strengthening this interpretation.  The procedure adopted by \cite{tak2017,tak2019} for sky subtraction is to fit a linear function to the sky brightness off disk, and then extrapolate that function to the locations on the disk at which they make their measurements.  In addition in some cases they add another linear term with a different slope to flatten the residuals in the lunar disk flux.  The limited region of sky flux in our images made this procedure difficult to implement.  We also had doubts that the sky flux would continue to increase as we moved further away from the lunar edge across the disk as we saw no evidence of an increase in raw flux in that direction.  Instead the variability appeared random.  Inspection of the images showed it was dominated by identifiable lunar surface features.

\begin{figure}
	\includegraphics[width=\columnwidth]{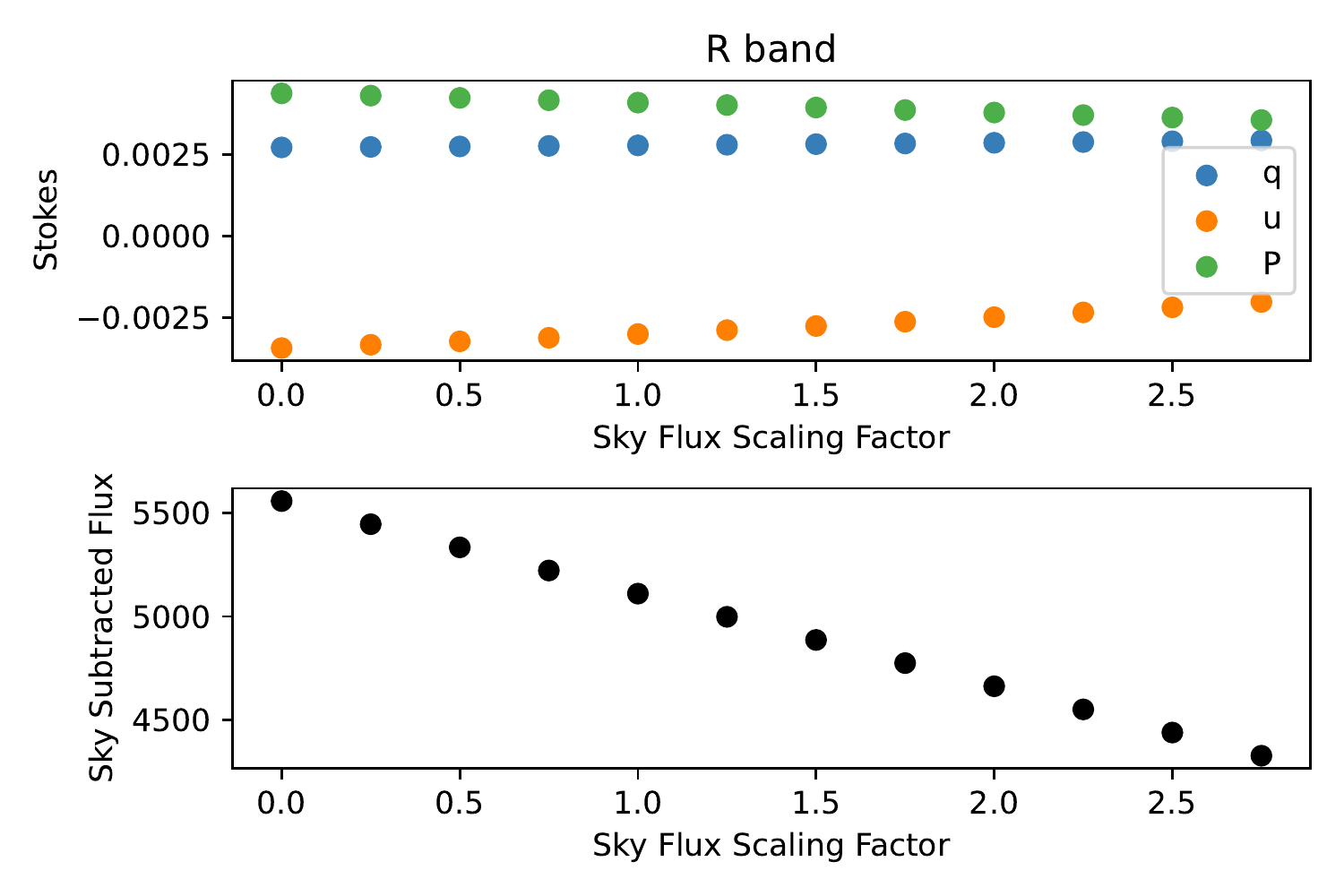}
	\includegraphics[width=\columnwidth]{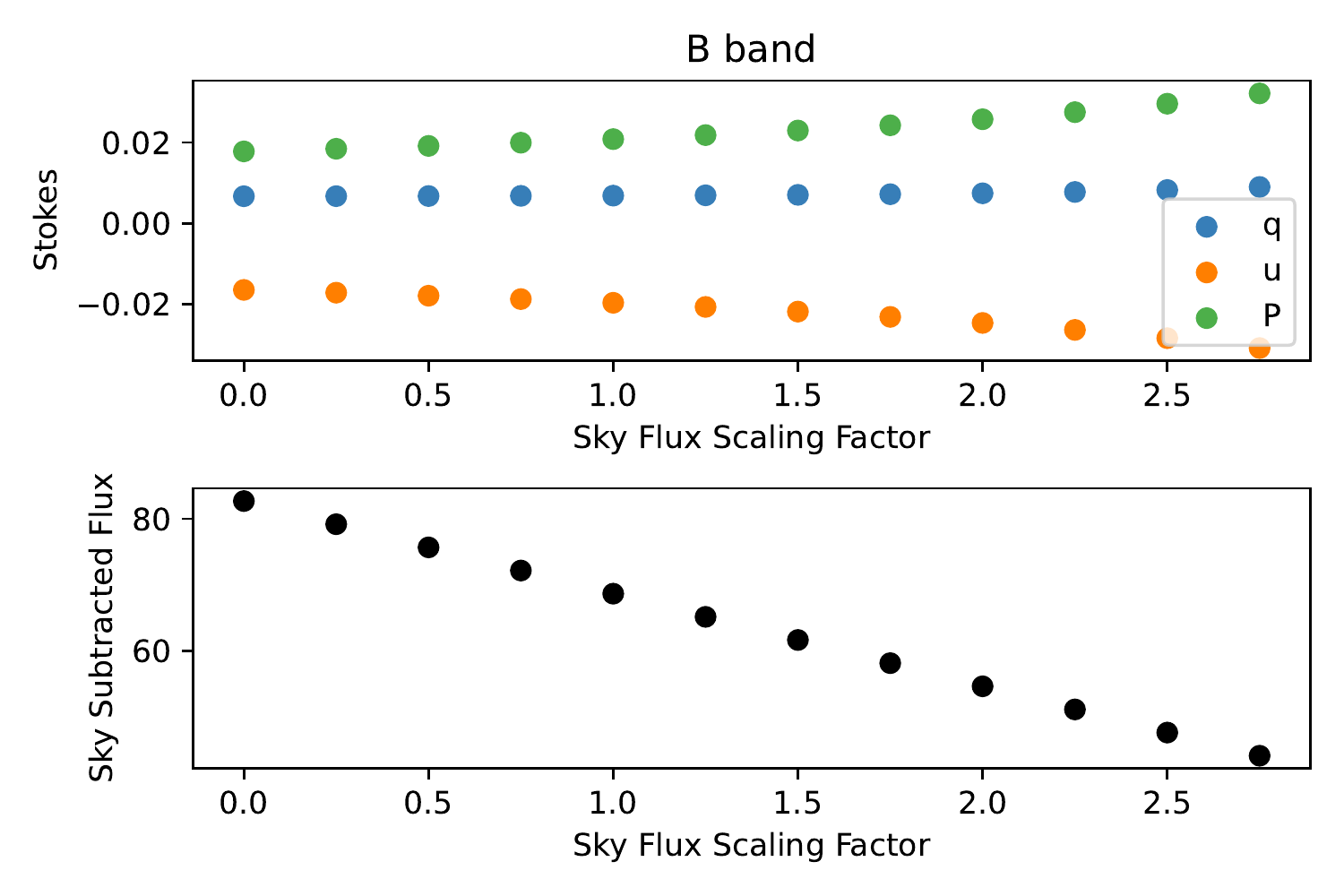}
    \caption{The effect of sky subtraction on measured Stokes parameters (not corrected for instrumental depolarization) in the $R$ (upper panels) and $B$ (lower panels). Flux is measured in units of ADU/pixel. Due to the very high count rates when the flux is integrated over the measuring areas, the formal (Poisson) errors are smaller than the plot symbols.}
    \label{fig:sky}
\end{figure}

This problem was particularly vexing, as due to the drift of the lunar disk over the detector, only the first two observing sequences listed in Table \ref{tab:log} (in the $R$ and $B$ bands) recorded areas outside of the lunar disk.
In order to evaluate the importance of sky subtraction, and explore if it was necessary for our data-set, we carried out an analysis on these first two sequences.

Firstly we defined a pair of sky regions of 30x100 pixels located $\sim30$ pixels from the lunar edge and used this on a per-frame basis to carry out sky subtraction from the measured mean counts in our the central regions located "above" them in the images (Fig. \ref{fig:moon-sky}).  Due to the telescope pointing drift ($0.54$-pixels/frame), the location of the defined regions were recalculated for each frame to keep the sky area and lunar features at the same effective positions.  The measured mean sky counts were $8$ per-cent of the counts from the lunar disk.

In order to understand the effect of the sky subtraction, the measured mean sky counts on a frame by frame basis were scaled by a scaling factor between 0.0 and 3.0 times, a factor of 0.0 corresponding to no sky subtraction and a factor of 3.0 corresponding to a very large sky subtraction ($25$ per-cent of the lunar disk value).  The scaled mean sky counts were then subtracted from their corresponding mean image region counts.  The impact of the various scaling factors can be seen in Fig. \ref{fig:sky} - upper panels, with a {\em decrease} in measured $R$-band polarization of $\sim 0.03$ per cent per unit scaling factor in a value of $0.43$ (i.e. a $7$ per cent relative decline). 

A similar analysis was carried out for the first $B$-band observing sequence (Fig \ref{fig:sky} - lower panels).  In this case, the measured mean sky counts were $12$ per-cent of the lunar disk value.  The polarization effect found was an {\em increase} in the measured value of $0.2$ per cent per unit scaling factor in a value of 1.77 per cent (i.e. a 11 per-cent relative increase).

Considering both of these results, we conclude overall that uncertainty in sky subtraction can introduce a relative percentage error in measured polarization values similar to the size of the ratio between the lunar disk and nearby sky. As a conservative limit, we therefore adopt a 20 per cent relative error in our final determinations to account for this affect.  By adopting this uncertainty, it gave us the ability to proceed with the analysis of all of our observing sequences (including the majority which were without blank sky regions) by only measuring the flux from the centre of the images and neglecting sky subtraction.  We note that this approach may only be valid during the total eclipse phase (as occurs in all of our data) and that during a partial eclipse the sky signal would be more intense and the gradient likely steeper. 

In support of this approach we note that measurements of the polarization of the sky subtraction areas themselves yield values of 1.0 percent in $B$ (EVPA=$33.5^\circ$) and 0.9 percent in $R$ (EVPA=$20.6^\circ$).  The similarity to the EVPA values of the lunar disk (Table \ref{tab:log}) indicates (as also found by \cite{tak2017}) that the likely origin of the sky polarization is  scattering by the Earth's atmosphere of polarized light from the Moon.  The importance of accurate sky subtraction for polarization measurements is therefore diminished.   Further support for this interpretation can be found by applying a simple model of sky polarization caused by scattered lunar light.  This method is outlined in Appendix A of \cite{gg} and predicts a change in foreground polarization due to scattering of $<0.001$ percent over the distance between between the sky region and the measurement areas.

\section{Results and Discussion}
\label{sec:results}

\subsection{Global Polarization Properties}

\begin{figure}
	\includegraphics[width=\columnwidth]{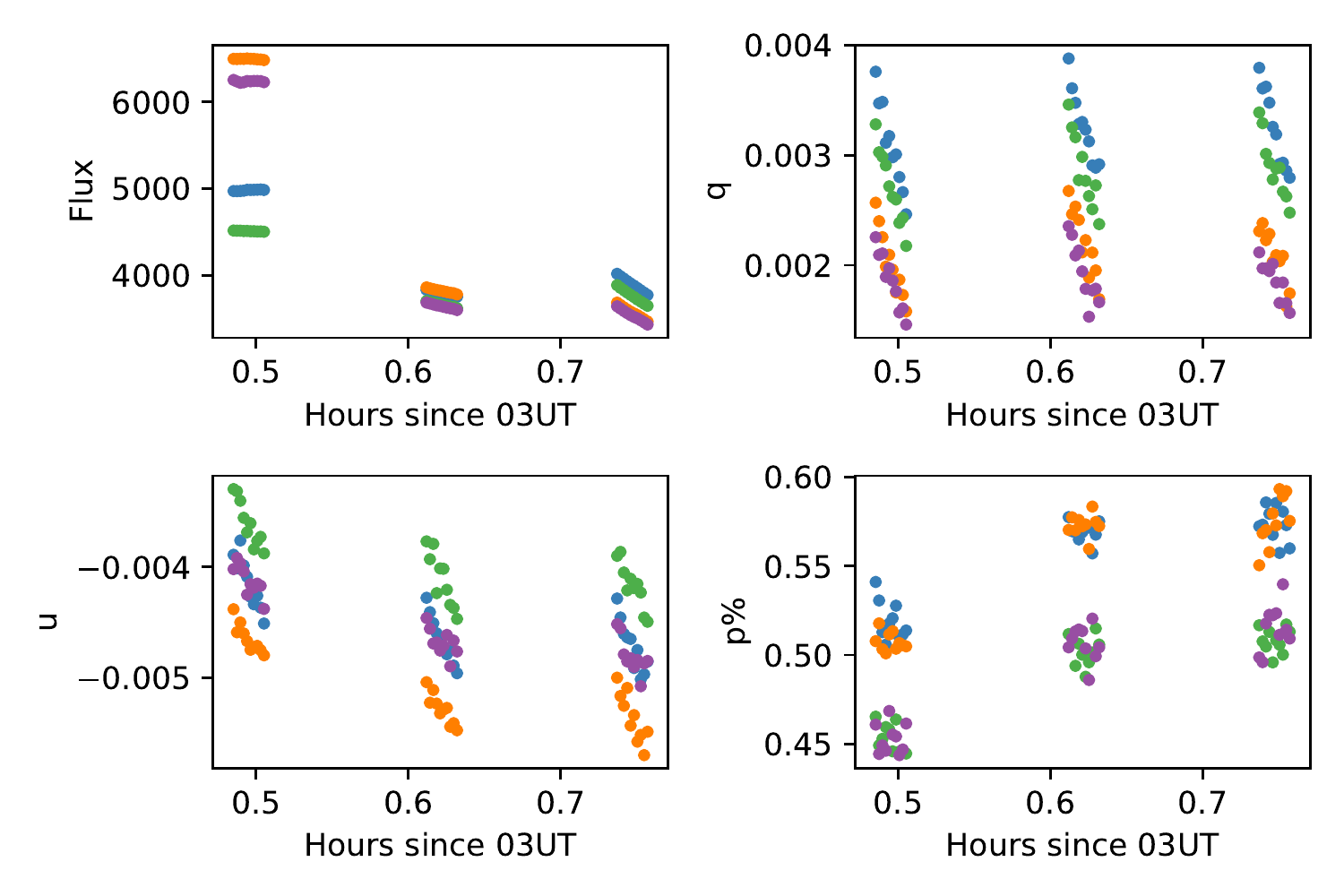}
	\includegraphics[width=\columnwidth]{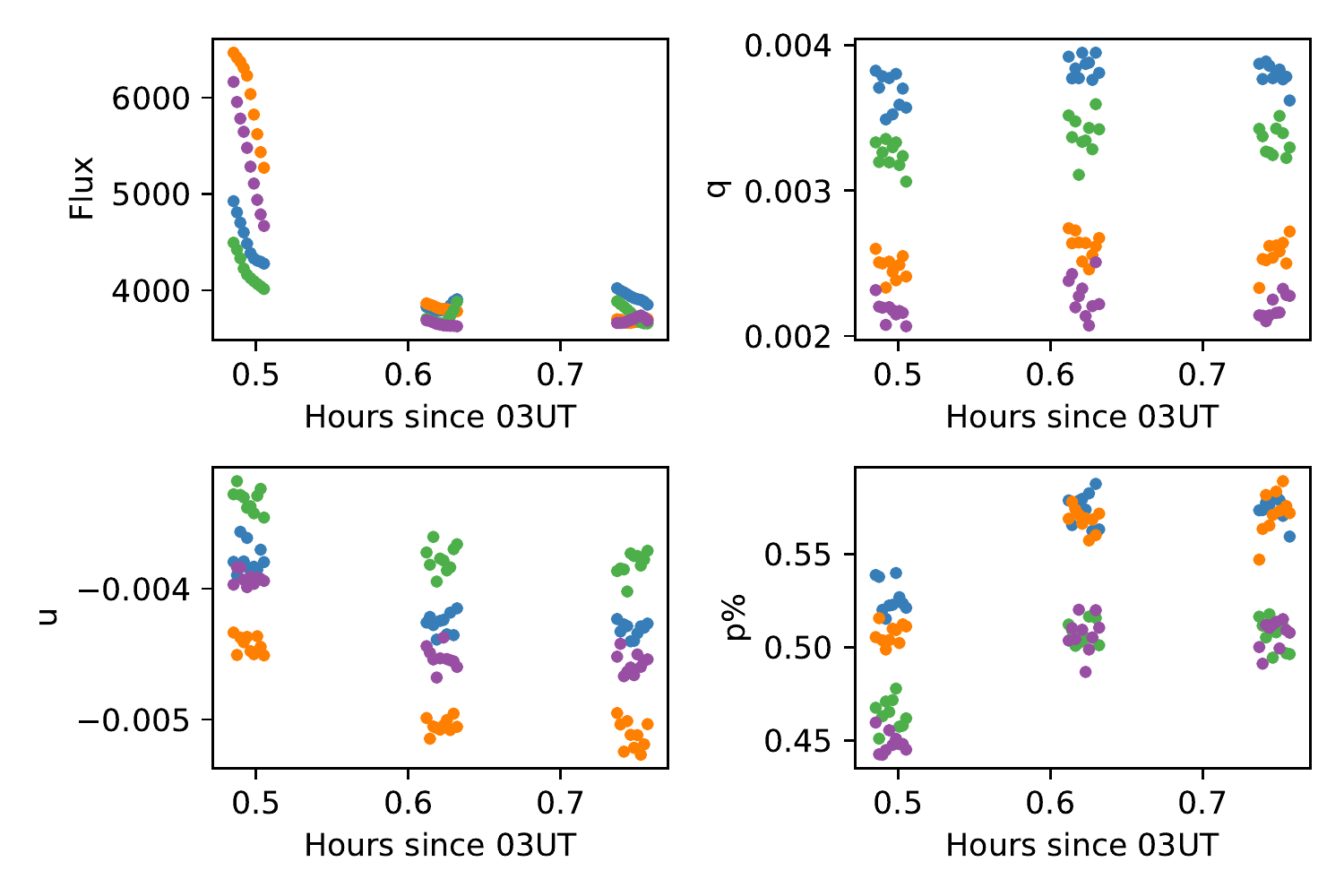}
    \caption{Measured Flux (ADU counts per pixel), $q$, $u$ and $P$ for the three $R$ band observing sequences. Formal error bars are smaller than the point sizes. The colours indicate the 4 measurement areas as shown in Fig \ref{fig:moon-sky}. The upper four panels are analysed with the location of the measuring areas tracking the drift of the lunar disk within each observing sequence.  The lower four panels keep the measuring areas the same for images.  While there are slight differences between the 4 measurement areas and the two approaches, the overall difference in the measured polarization is negligible. There is slight evidence for an increase in polarization between the first and subsequent observing sequences (see text for discussion).
    }
    \label{fig:drift}
\end{figure}

Based on the analysis presented in Section \ref{sec:sky} we decided for consistency to derive our final results for all observing sequences without sky subtraction.  The only remaining question was whether or not to track our four central measuring areas (Fig. \ref{fig:moon-long})) across an observing sequences to remove the effect of the non sidereal motion.  We evaluated both approaches and the results are presented in Fig. \ref{fig:drift}. In the figure the parameters are presented separately for the 4 measuring regions in different colours.  In both cases a repeatable small systematic offset between the different regions of about 0.1 per cent is visible and is an indication of the presence of the slight image position dependant instrumental polarization.  These offsets were of very similar magnitude and spatial distribution in all filters and observing sequences and give a $\pm0.05$ per cent error contribution to the final polarization measurements.  A very slight increase in polarization ($\sim0.05$ percent) may be visible between the first and second observing sequences in $R$ if one assumes the 20 percent sky subtraction error in polarization is systematically identical between sequences in a given filter.  The start of total eclipse began simultaneously with the start of the first observing sequence, however the photometry shows the lunar disk was $\sim0.5$ magnitudes brighter at this time than during the second $R$ band sequence and it may be that an implied difference in atmospheric transmission from the Sun to the Moon via the Earth's atmosphere could explain a slight polarization change.

From Fig. \ref{fig:drift} it is also apparent that tracking the lunar surface makes the flux values more stable (as would be expected since this is a measurement of visible surface features) but slightly increases the scatter on a set of polarization measurements within a measuring area.  Ultimately the differences in measured polarization are not significant, but we chose to use the non tracking approach as our goal was polarization measurement.

Our final polarization measurements for all filters and sequences are presented in Table \ref{tab:log}. Error values on all measurements are based on combining the errors from the sky subtraction, the location dependant instrumental polarization and the instrumental polarization zero-points (Table \ref{tab:filters}).  The absolute measurements within each band are consistent to $<0.1$ per-cent with no indication of time variability over the $\sim15$ minute observing run.  As noted above there is slight evidence of an increase of $\sim 0.05$ percent in the relative values between the first two $R$ band observing sequences when the lunar disk was still dimming (although technically fully eclipsed) although this interpretation relies on an assumption of consistency in sky subtraction between the runs and should not be over-relied upon.

The mean percentage polarizations are observed to decrease monotonically with wavelength:  $2.1\pm0.4$ in $B$, $1.2\pm0.3$ in $V$, $0.5\pm0.2$ in $R$ and $0.2\pm0.2$ in $I$.  No correction for depolarization by back-scattering at the lunar surface has been applied to these values.  From a compilation of measurements of lunar samples reported in the literature \cite{bazzon} estimate this may decrease the measured values by up to two-thirds compared to the incident values.  The true polarization in the $B$-band of the incident light could therefore be as high as $\sim6$ percent.

The $B$- and $V$-band detections of polarization appear strongly statistically significant, while the $R$ band detections are somewhat marginal and the $I$ band definitely not significant.  There is good agreement ($\pm 1^\circ$) between the EVPA values in the $B$, $V$ and $R$ bands, further increasing our confidence in these detections.  The $I$ band EVPAs sit well away from the other bands, and are further evidence of the non-significant nature of the polarization measurement in that band.  As far as we are aware this is the first $B$ band measurement of the eclipsed Moon.  The high value indicates that future observing programmes should also consider including $B$ in their measurement strategy. Our $I$ band limit is consistent with the previous data from both \cite{tak2017} and \cite{strassmeier} who only found limited polarization around the O2 band. 

The most interesting result is that for the $V$ band due to our ability to compare with previous observations.  As mentioned in the introduction, the few previous measurements of the eclipsed Moon at this wavelength are between $\sim 2.5$ and $<1$ percent.  Our value of $1.2\pm0.3$ percent sits in the middle of the range of previous measurements, and adds further evidence for the long term time variability of this quantity.  A variety of possible explanations for this long term variability have been proposed by \cite{tak2017, tak2019} who made a detailed comparison of Earth observation and meteorological records of the Earth's atmosphere at the times of the two eclipses they observed.    

The colours and photometric depths of lunar eclipses can vary dramatically with a proposed correlation with the presence of ash and other particulates high in the Earth's atmosphere \citep{strothers,strothers2,garcia}.  It is plausible that this could have some effect on the degree of polarization observed.    

Comparison of our photometry (Table \ref{tab:log}) with the computed value from the JPL Horizons database shows an eclipse depth $\Delta V \sim 12.2$.  Unfortunately no reliable photometric measurements of previous eclipses with polarimetric data could be found in searches of the literature.  However values of between $\Delta V\sim10$ \citep{schober} and $\Delta V\sim 16$ \citep{mat-dec64} have been reported in the literature for other total eclipses, indicating this eclipse lies in the middle of the possible range of values.  Similarly during the eclipse we measured $B-V=1.6\pm0.1$.  Typical $B-V$ values during eclipses vary from $B-V\sim0.8$ \citep{mat-dec63} to $B-V\sim2.4$ \citep{mat-dec64}, again placing our measured value in the middle of the possible range.  Overall it therefore appears that the middling photometric and polarimetric properties of the May 2022 eclipse may represent those to be expected in somewhat typical atmospheric conditions.  

\subsection{Local Polarization Properties}

\begin{table}
	\begin{center}
	\caption{List of Observing Sequences and final degree of polarization ($P$) measurements.  These values have been corrected for instrumental polarization and depolarization.  Polarization bias corrections \citep{bias} have been applied but are negligible.  Electric Vector Polarization Angle (EVPA) is measured increasing East of North in the equatorial (sky) coordinate system as per the standard convention \citep{evpa}.  SB is the surface brightness in magnitudes per square arcsecond.}
	\label{tab:log}
	\begin{tabular}{ccccc}
		\hline
		Time  & Filter &  $P$ & EVPA  & SB \\
		(UTC) & & \% & $^\circ (\pm1^\circ)$  & mag/sq. arcsec\\
		\hline
		03:29:02 -- 03:30:22 & $R$ & $0.5\pm0.2$ & 26.6 & $12.84 \pm 0.15$ \\ 
		03:30:55 -- 03:32:15 & $B$ & $2.1\pm0.4$ & 25.5 & $17.19 \pm 0.08$\\
		03:32:51 -- 03:34:11 & $V$ & $1.2\pm0.3$ & 27.2 & $15.70 \pm 0.04$\\
		03:34:45 -- 03:36:04 & $I$ & $0.2\pm0.2$ & $105.4^*$ &$11.01 \pm 0.02$\\ \\
		03:36:38 -- 03:37:58 & $R$ & $0.5\pm0.2$ & 25.7 & $13.27 \pm 0.02$ \\
		03:38:28 -- 03:39:50 & $B$ & $2.1\pm0.4$ & 26.7 & $17.45 \pm 0.08$\\
		03:40:26 -- 03:41:46 & $V$ & $1.2\pm0.3$ & 28.6 & $15.73 \pm 0.05$\\
		03:42:16 -- 03:43:36 & $I$ & $0.2\pm0.2$ & $98.4^*$ & $11.01 \pm 0.04$\\
		\\
		03:44:09 -- 03:45:29 & $R$ & $0.5\pm0.2$ & 25.0 &  $13.28 \pm 0.04$ \\
		\hline
	\end{tabular}
	\end{center}
	$^*$EVPA values in the $I$ band are meaningless due to the non-detection of polarization.
\end{table}

It is possible that the differing surface properties of the different geological areas of the Moon may cause a change in the degree of depolarization suffered by the incoming polarized light when it is reflected back towards the Earth.
\cite{bazzon} show a correlation between albedo and degree of depolarization caused by back-scattering of lunar samples measured by \cite{hapke}.  The correlation would predict (their Fig 11) a change in polarization efficiency from around 0.50 to 0.45 between dark and bright regions.  This would correspond to a predicted relative increase in measured polarization by around 10 percent in the dark regions over the lighter ones.

To investigate this we made a brief analysis of the properties of the $B$-band polarization comparing two geologically different areas of the Moon. For this analysis the effects of sky calibration uncertainty can be neglected as we are only interested if any polarization difference exists.  We made our analysis on the first $B$-band sequence given the higher polarization at this wavelength and the presence of well defined Mare and brighter regions.  We tracked a pair of $100\times60$ pixel extraction regions across the observing sequence.  The first region covered the dark Mare Grimaldi and the second the directly adjacent brighter upland region.  The flux difference between the brighter and darker regions was $\sim 25$ percent.  Grimaldi gave a polarization value of $2.07\pm0.02$ percent and the adjacent brighter region $2.05\pm0.02$ percent.  No significant difference between the two regions was therefore apparent.  While this 
strengthens our general interpretation of the origin of the polarization signal as being external to the Moon, it is in conflict with the prediction discussed in the previous paragraph.  A more detailed analysis could be conducted by constructing detailed spatially resolved polarization maps to try and resolve this tension.  However the significant position dependant instrumental polarization in MOPTOP made this too challenging to attempt with this particular dataset.

Similarly we made a comparison of the polarization properties the near the limb of the Moon and closer to the centre of the lunar disk.  To avoid the uncertainty introduced by the effect of the large scale spatially dependant instrumental polarization in the instrument, we compared values for the same detector area (a region of $100\times50$ pixels) between observing sequences with matching filters, relying on the drift of the Moon to investigate the affect of angles of incidence (AOI).  In the $B$-band we measured polarization of $1.76\pm0.10$ percent for observations near ($\sim20$ arcsec, ${\rm AOI}=76^\circ$) the limb and $1.74\pm0.10$ for observations $\sim 3$ arcmin (${\rm AOI}=55^\circ$) from the limb\footnote{The lower polarization values here compared to those quoted for our other measurements reflect the different instrumental polarization characteristics near the edge of the frames and mean these values can not be compared with others in this work.}.  No significant difference in polarization between the different illumination angles is apparent.

\section{Conclusions}
\label{sec:conc}

We have presented multi-band polarimetry of the lunar disk for the first 15 minutes of total lunar eclipse on 2022 May 16.  A lack of non-sidereal tracking meant the lunar image drifted over the instrument focal plane, and limited our ability to carry out sky subtraction.  We have shown this introduces an relative error of up to 20 percent into our measurements.    Our final percentage polarization values are $2.1\pm0.4$ in $B$, $1.2\pm0.3$ in $V$, $0.5\pm0.2$ in $R$ and $0.2\pm0.2$ in $I$.  These values lie in-between those observed in the few previous measurements made, and provide support for the analysis presented in \cite{tak2019} that the eclipsed lunar polarization may be time variable between eclipses.  No strong evidence was found for short term (seconds to minutes) variability.  Considering the measured eclipse photometric depth and colour the eclipse seems to have been fairly typical, and the measured polarization values may therefore reflect the average properties of the eclipsed moon.   We also found no significant polarization difference ($<0.02$ percent) between a region of dark Mare and nearby bright uplands or between the lunar limb and regions closer to the disk centre (i.e. due to the differing angle of incidence).  This further strengthens the interpretation of this affect as one due to scattering in the Earth's atmosphere rather than by the lunar regolith.

The next total lunar eclipse on 2022 Nov 8 will not be visible from La Palma. However the following two eclipses of 2025 March 14 and 2025 Sept 8 will be observable from that location.  By implementing a regular re-pointing strategy during observations to ensure better sky sampling, we aim to improve our measurement error to $\pm0.1$ for those events.  In addition we note a number of other possible optimizations for future observing programmes on all telescopes:
\begin{enumerate}
    \item While this observation set was truncated by the telescope altitude limit, in general the opportunity to observe for a period of several hours through the entirety of totality and into the penumbral phase should be taken.  This would allow an investigation of longer timescale polarization variability as the region of atmosphere responsible was changed by the Earth's rotation.  In addition the opportunity should be taken to optimize the observing cadences between filters and locations on the lunar disk to probe all potential variability timescales.
    \item Optimization of the filters used.  From our observations it appears that $B$ and $V$ are key.  However an extension into the $U$-band would presumably further increase our sensitivity to scattering induced polarization.
    \item The opportunity to collect and analyse photometric data at the same time as the polarimetric data should be taken to allow correlation with eclipse depth and colour.
    \item It may also be interesting to consider attempting to observe circular polarization.
    \item A strategy of deliberate observation tiling with appropriate overlaps and offsets to nearby sky would allow proper mapping of off-axis instrumental polarization and improve the ability to investigate any effects due to lunar geology, the effects of reflection angles, and the distance from the umbra centre.
\end{enumerate}

\section*{Acknowledgements}

  The Liverpool Telescope is operated on the island of La Palma by Liverpool John Moores University in the Spanish Observatorio del Roque de los Muchachos of the Instituto de Astrofisica de Canarias with financial support from the UK Science and Technology Facilities Council under UKRI grant ST/T00147X/1. This research made use of Astropy,\footnote{https://www.astropy.org} a community-developed core Python package for Astronomy \citep{astropy13, astropy18}. 
KW acknowledges support through a UK Research and Innovation Future Leaders Fellowship awarded to dr.~B.~Simmons (MR/T044136/1). This work was supported by Towards Turing 2.0 under the EPSRC Grant EP/W037211/1 \& The Alan Turing Institute: KW acknowledges support through an Alan Turing Institute Post-Doctoral Enrichment Award, this project was formulated during a visit funded by this award.  KW thanks R. Starling, B. Simmons and P. Chote for useful discussions. IAS similarly thanks J. M. Steele.  In the planning of these observations, we made use of the Stellarium software (\url{http://stellarium.org}) and extensive use of the book The Moon by \cite{Moonbook}; KW is grateful to the late Prof.~Pik-Sin Th\'{e} for donating him a copy of this book.

\section*{Data Availability}

 All data is available to download from the Liverpool Telescope data archive\footnote{https://telescope.livjm.ac.uk/}.



\bibliographystyle{mnras}
\bibliography{example} 







\bsp	
\label{lastpage}
\end{document}